\documentclass[letterpaper,11pt]{article}
\usepackage[utf8]{inputenc}
\usepackage{amsmath, amsthm, amssymb, amsfonts}
\usepackage{hyperref}
\usepackage{graphicx}
\usepackage[normalem]{ulem}
\usepackage{cite}
\usepackage[font=footnotesize]{caption}
\usepackage{algorithm}
\usepackage[noend]{algpseudocode}
\usepackage{authblk}
\usepackage{float}
\newcommand{\var}{\texttt}
\newcommand\tab[1][6mm]{\hspace*{#1}}
\newenvironment{mat}{\left[\begin{array}{ccccccccccccccc}}{\end{array}\right]}
\newcommand\bcm{\begin{mat}}
\newcommand\ecm{\end{mat}}

\makeatletter
\def\BState{\State\hskip-\ALG@thistlm}
\makeatother

\title{Functional Connectomics from Data: Probabilistic Graphical Models for Neuronal Network of C. elegans}

\author[1]{Hexuan Liu}
\author[2]{Jimin Kim}
\author[1,2]{Eli Shlizerman}

\affil[1]{Department of Applied Mathematics, University of Washington, 98195, USA}
\affil[2]{Department of Electrical Engineering, University of Washington, 98195, USA}

\date{\today}

\begin{document}

\maketitle

\textbf{Abstract} 
We propose a data-driven approach to represent neuronal network dynamics as a Probabilistic Graphical Model (PGM). Our approach learns the PGM structure by employing dimension reduction to network response dynamics evoked by stimuli applied to each neuron separately. The outcome model captures how stimuli propagate through the network and thus represents functional dependencies between neurons, i.e., functional connectome.
The benefit of using a PGM as the functional connectome is that posterior inference can be done efficiently and circumvent the complexities in direct inference of response pathways in dynamic neuronal networks. In particular, posterior inference reveals the relations between known stimuli and downstream neurons, or allows to query which stimuli are associated with downstream neurons. For validation and as an example for our approach we apply our methodology to a model of \textit{Caenorhabiditis elegans} nervous system which structure and dynamics are well-studied. From its dynamical model we collect time series of the network response, and use singular value decomposition to obtain a low-dimensional projection of the time series data. We then extract dominant patterns in each data matrix to get pairwise dependency information, and create a graphical model for the full somatic nervous system. The PGM enables us to obtain and verify underlying neuronal pathways dominant for known behavioral scenarios, and to detect possible pathways for novel scenarios.

\section{Introduction}
\tab Probabilistic Graphical Model (PGM) is a statistical model in which a graph maps the conditional dependence structure between random variables \cite{nutshell,Koller,Murphy}. A complete graphical model contains structural relations, as well as parameters that define the conditional dependencies between the random variables. Most commonly used graphical models are Bayesian networks and Markov random fields. Bayesian networks are directed graphs parametrized by conditional probability distributions (CPDs), whereas Markov random fields are undirected graphs parametrized by factors. There are two types of inference problems in the context of network pathways that can be answered using a graphical model. The first problem is to infer conditional probabilities of the type $P(Y|E=e)=\frac{P(Y,e)}{P(e)}$, where the probability distribution over the values $y \in Y$, conditioned on $E=e$ is inferred. For example, in such a case pathways from upstream neurons belonging to the set $E$ are sought. The second problem is to infer the maximum a posteriori (MAP) probability: $\arg\max_y P(y|e)$, where the most likely assignment to the variables in $Y$ is sought, given the evidence $E=e$. In this problem, dominant pathways leading to a set of downstream neurons and revealing upstream neurons activating them are sought. PGM is capable of discovering the unstructured information within the distributions, since it turns complex distributions into structured information that can be analyzed effectively and computationally efficiently.

PGMs were successfully applied to problems for which direct inference is intractable and therefore appeal to apply in the context of functionality of neuronal networks \cite{Koller}.  However, classical approaches for learning PGM structure are designed for discrete variables and are not compatible with neuronal networks consisting of dynamic neurons interacting with each other through dynamic connections. Both neurons and their connections are typically modeled as nonlinear processes. Possible adaptation of neuronal network to a statistical model which captures functionality is to consider each neuron as a random variable $X_i$ representing the states of the neuron from the set $\mathcal{X}$ containing all neural state variables. For simplicity, it is often assumed, and here we assume it as well, that each neuron activity is binary-valued, i.e., $X_i\in \{0,1\}$, with 0 being the inactive state and 1 being the active state. The PGM is thereby a graph with $X_i$ being nodes and dependencies as edges between the nodes.

There are two main difficulties in learning a graphical model from dynamics: (i) Difficulty in estimating input-output correlations for a network which responses are time dependent. (ii) High dimensionality and complexity of neuronal networks often incorporating loops. These factors make statistical inference hard to realize. Relevant work has been done in dynamic networks using either experimental data, such as in \cite{Bayesian,varying1, varying2}, or simulated data, such as in \cite{TE}. In both cases, ``snapshots" that record  network dynamics are needed to analyze time-series data. To estimate input-output correlations, statistical methods are applied to measure pairwise node correlation. For example, Honey et al. proposed to use transfer entropy (TE) to capture patterns of directed interaction and information flow between pairs of nodes \cite{TE}. Butte and Kohane used entropy of gene expression patterns and the mutual information between RNA expression patterns for each pair of genes to compute pairwise mutual information \cite{Entropy}. Existing works often assume that the underlying network is time-invariant, such as \cite{Bayesian,Basso}. Work has also been done towards estimating a sequence of graph structures that capture the network rewiring processes. \cite{varying1,varying2}. Most of the existing learning methods sample the neuronal network multiple times as training data to evaluate candidate models, and then search for the optimal model according to a scoring function. They fall into the category of "score-based approaches" as described in \cite{nutshell}. One of the drawbacks of score-based approaches is that in neuronal networks with high dimensional state space and cyclic structure, the problem to find an optimal solution is often NP-hard. Here we propose a new approach that could possibly circumvent these complexities. Instead of learning the graph structure and parameters from sampled data, we learn from low-dimensional projections of time-series that capture the full dynamics of the network, and thus reduce the complexity of the learning algorithm and the computational cost.

Our approach is to simulate the network to obtain time-series data of neural activity, and to estimate the input-output correlations by projecting the data to lower dimensions. Our key idea is that we stimulate each single neuron independently, i.e., orthogonal stimulation, and record network response to each such stimulus. We, therefore, inherently assume that each neuron stimulation is independent and the PGM superimposes the response dynamics to these independent experiments into a model. The projected data contains dominant patterns of network dynamics over time which we translate to pairwise probabilistic dependencies between neurons. To be able to capture the dominant patterns we sample the network over a sufficiently long period of time. Using this approach, we construct a PGM, specifically a Bayesian network, mapping the functional connectivity between the neurons. The model can be used for inference of marginal probabilities or MAP paths in the network.
\begin{figure}[t]
  \[\includegraphics[width=14cm]{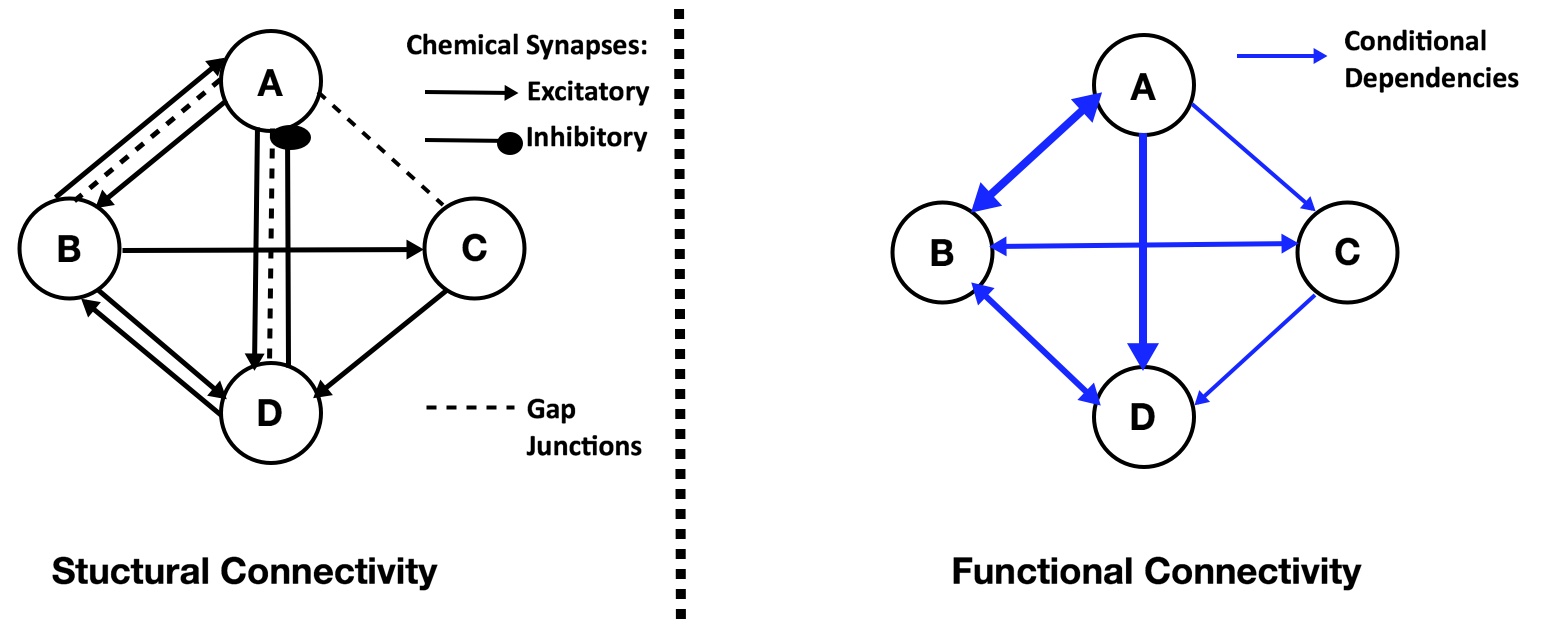}\]
  \caption{\textbf{Representation of a neuronal network as a graphical model.} Left: Example of structural/ anatomical connectivity map in which nodes denote neurons and edges map connections, e.g., chemical synapses and electrical gap junctions. Interactions between neurons produce a nonlinear network which dynamically transports stimuli to neuronal behaviors. Right: Example of PGM constructed from the neuronal network governed by the structural connectivity map and nonlinear dynamics. In the PGM nodes are random variables corresponding to neuronal states and edges are conditional probabilities. PGM structure captures functionality of the network and hence typically different from the anatomical connectivity map.}
\end{figure}

We apply our method to the neuronal network of Caenorhabiditis elegans (\textit{C. elegans}), a well-studied organism, to validate the PGM. The somatic nervous system of \textit{C. elegans} consists of  279 neurons and the wiring diagram between these neurons was compiled by Varshney et al. consisting of 6393 chemical synapses, 890 gap junctions, and 1410 neuro-muscular junctions \cite{connectome}. The connectome represents weights of  dynamic interactions between neurons and combined with a biophysical model of neural dynamics and their interactions enables us to simulate the full nervous system model \cite{dynamic1,dynamic2,dynome}. In the following sections we construct a Bayesian network that represents the functional connectivity of the neuronal network of \textit{C. elegans} and verify the results with experimental data.

\section{Construction of Probabilistic Graphical Model from Neuronal Network Dynamics}\label{sec:construction}
\tab We first introduce the components of Bayesian network in the context of neuronal networks. We then illustrate how conditional probabilities are assigned from the simulated neural dynamics. Finally, we show how computed probabilistic dependencies are used for construction of a Bayesian network.    

\subsection{Representing Neuronal Distributions with Bayesian Networks}
\tab A Bayesian network is a representation of a joint probability distribution with graph structure $G=(V, E)$ and parameters $\theta$. The graph structure $G$ is a directed acyclic graph (DAG) whose vertices, $V$, correspond to random variables $\{X_1, X_2, \cdots X_n\}$, and edges, $E$, correspond to connections between random variables, i.e., conditional dependencies. The parameter $\theta$ describes the conditional probability distribution for each variable given its parents in $G$. By applying the probability chain rule and using properties of conditional independence, joint probability distribution can be decomposed into the product form
\[P(X_1, \cdots, X_n)=\Pi_{i=1}^n P(X_i|\textbf{Pa}^G(X_i))\]
where $\textbf{Pa}^G(X_i)$ is the set of parents of $X_i$ in $G$.\\

PGM is based on this property and graphically reformats probability distributions into a graph with parent and children nodes. For neuronal networks, each $X_i$ corresponds to individual neuron where each of them takes discrete values. Thus we can represent $P(X_i|U_1, \cdots, U_k)$ as a table that specifies the probability of values for a neuron $X_i$ for each joint assignment to its parent neurons $U_1, \cdots, U_k$ \cite{Bayesian}. In the simplest case we only consider pairwise conditional probability $P(X_i=s_l|X_j=s_k)$, where neuron $X_j$ receives input that drives it to some neural state $s_k$ out of set of states $s=\{s_1,..,s_k,..,s_m\}$, and we would like to estimate the probability of neuron $X_i$ being driven into state $s_l$. 

\subsection{Collecting Data from Simulated Dynamics}
\tab We propose to find conditional probabilities for all pairs of neurons by simulating the network and recording data in snapshot matrices. The snapshot matrix represents finite time series of whole network dynamics for a particular input to a single neuron. The matrices are computed by injecting a constant input into each and every neuron in the network independently, thus resulting in total $n$ matrices for the network of $n$ neurons. Network dynamics are modeled by a set of dynamic equations, e.g., conductance based model, which describe the biophysical processes of neurons and interactions between neurons. A snapshot matrix has the structure $S=[S(t_0)\ S(t_1) \ \cdots S(T)]$ of dimension $n \times T$, where $n$ is the total number of neurons in the network, and $t_0$ and $T$ is are start and end times of the simulation respectively. We assume that the sequence of random variables is independent and identically distributed (iid).

\subsection{Dimension Reduction}
\tab We process the time series data by projecting it into a lower dimensional space using singular value decomposition (SVD) to reduce computational complexity,. SVD of a $n \times p$ matrix $S$ has the form
\[S=U\cdot \Sigma \cdot V^*\]
Here $U$ and $V$ are $n\times n$ and $p \times p$ unitary matrices, with the columns of $U$ spanning the column space of $X$ and the columns of $V$ spanning the row space \cite{ESL}. $\Sigma$ is assumed to have its diagonal entries $\sigma_j$ which are non-negative and in descending order, i.e., $\sigma_1\geq \sigma_2 \geq \cdots \geq \sigma_m\geq 0$, where $m=\min(n,p)$. Once the original matrix $S$ is decomposed, $U_j$ (row vector of $U$) and $V_j$ (row vector of $V$) record the time dependent coefficients and the mode corresponding to singular value $\sigma_j$. Each singular value $\sigma_j$ in $\Sigma$ corresponds to the significance of the corresponding mode. Using SVD we can find a lower dimensional representation for the matrix $S$, which captures the dominant features of neural dynamics. Dimension reduction is performed by retaining $k-$dimensional subspace, where $k<n$, spanned by $k$-modes corresponding to top $k$ singular values.

When using SVD to obtain a low-dimensional representation of the data, we need to retain enough modes to approximate the data. In practice, energy criterion on singular values which specifies the amount of energy included in the chosen modes is often used. For example, typical energy criterion is to require 95\% of the energy in $\Sigma$. That is, the sum of the squares of the retained singular values should be at least 95\% of the sum of the squares of all singular values. For snapshot matrix which network responses reach a fixed point, the first dominant mode would be sufficient to represent the fixed point. For oscillatory networks, the first two modes together represent oscillatory dynamics. Instead of using the classical $k$-rank approximation, we propose a new approach that takes the linear combination of modes according to their significance, which results in a one-dimensional column vector. Specifically, we use the sum of all modes weighted by singular values as a one-dimensional representation of the original $\mathbb{R}^{n\times p}$ data
\[\sum_{i=1}^n \frac{\sigma_i^2}{\sum_{j=1}^n \sigma_j^2} |U_i|\in\mathbb{R}^n\].
\begin{figure}[h!]
  \[\includegraphics[width=\textwidth]{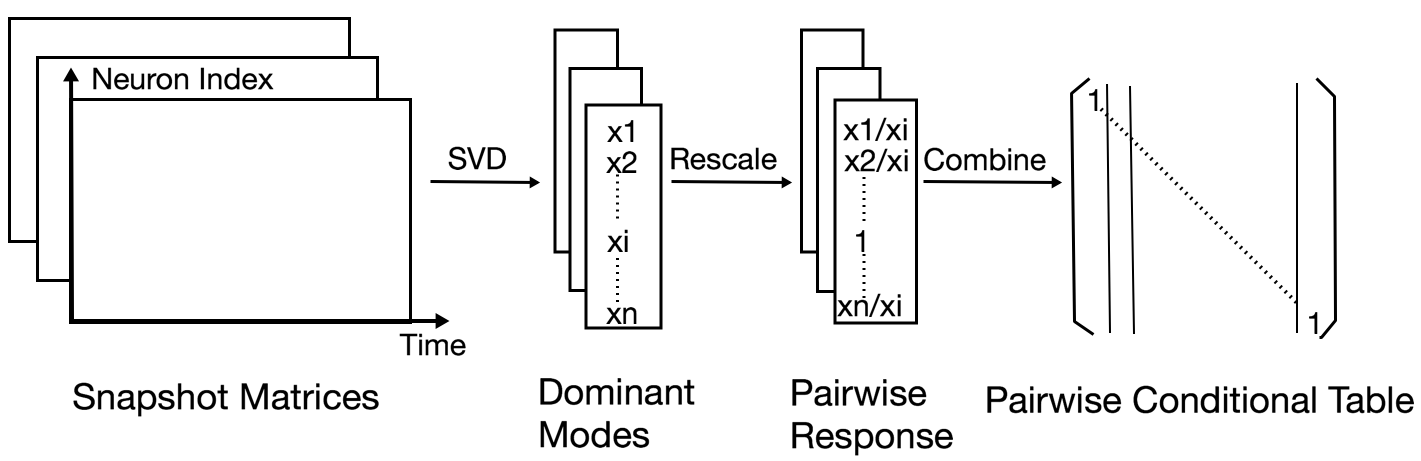}\]
  \caption{\textbf{Construction of dependencies between nodes.} By injecting input stimuli into each neuron, we obtain a series of snapshot matrices that record network dynamics. By decomposing the snapshot matrices using SVD we obtain the decomposition modes and compress them into a single vector for each matrix. We then normalize the vector according to its input neuron and get pairwise responses, which constitute the adjacency matrix of pairwise conditional probabilities.}
\end{figure}

\begin{algorithm}
\caption{Constructing Dependencies}\label{prob}
\begin{algorithmic}[1]
\State $n \gets \text{number of }\textit{neurons}$
\State $t_0 \gets \textit{starting time}$
\State $T \gets \textit{final time}$
\State $k \gets \textit{time step}$
\For{$i \in (0,n)$ } 
\State {$\text{Input}[i]=1$}
\State $X=run\_network(t_0, T,k,\text{Input})$
\State $[U,S,V]=SVD(X)$
\State $U_1=U[:,0]$
\State $prob=abs(U_1)/abs(U_1)[i]$
\EndFor
\end{algorithmic}
\end{algorithm}

\begin{figure}[h!]
  \[\includegraphics[width=12cm]{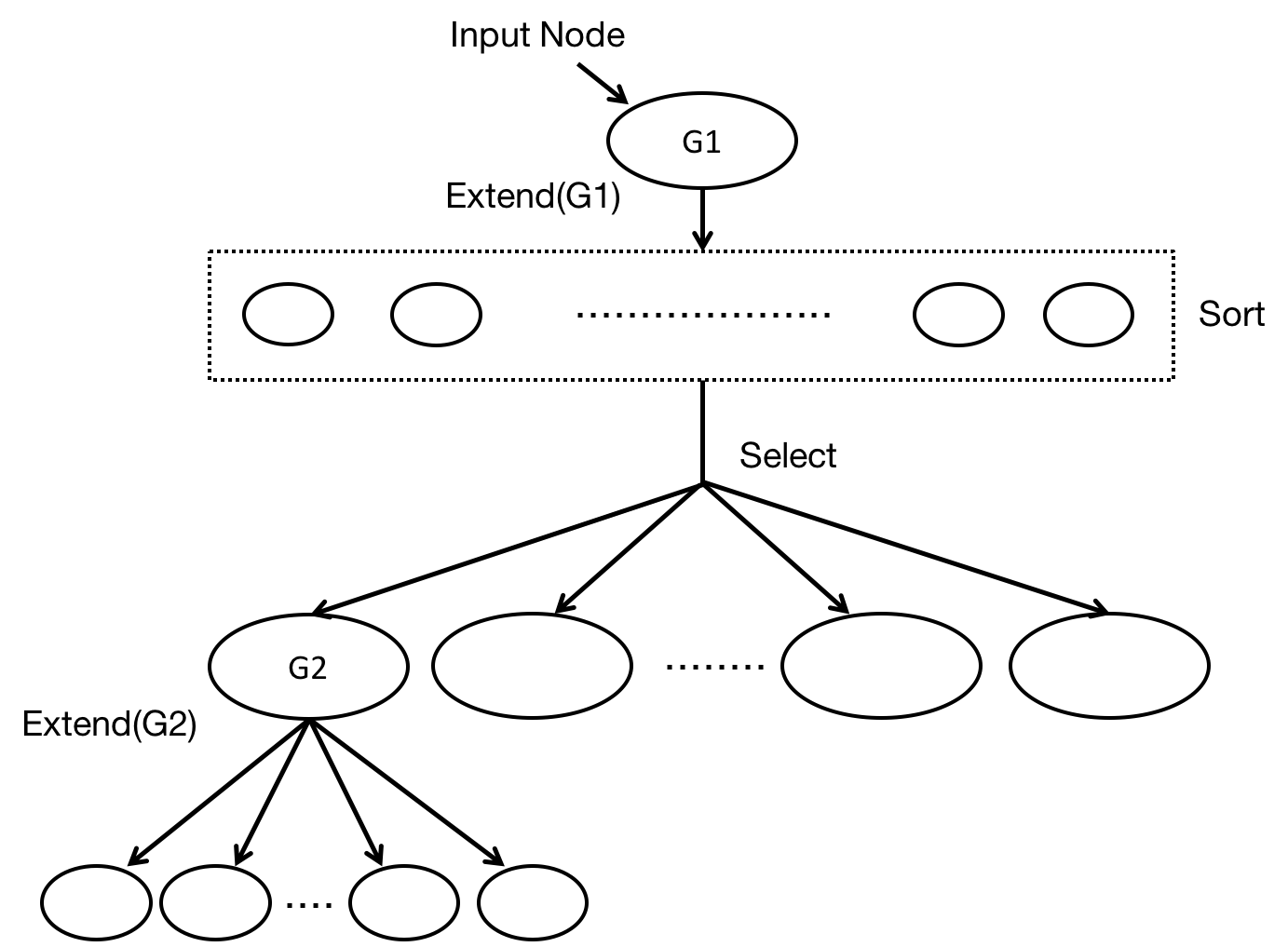}\]
  \caption{\textbf{Tree construction from adjacency matrices.} We start with the input node, and sort the other nodes according to their probabilities conditioned on the input node. We ignore nodes with conditional probabilities lower than the threshold, and select the top nodes with the highest conditional probabilities. A constrain can be imposed to limit the number of children that a node can have. From there we extend each node recursively, until it either reaches the maximum tree depth, or reaches neurons in a particular set (e.g. motor neurons). The threshold, maximum tree depth and maximum number of children are pre-set parameters to limit the size of the tree. }
\end{figure}
\begin{algorithm}
\caption{Constructing Trees}\label{tree}
\begin{algorithmic}[2]
\State {$\var{maxLayer} \leftarrow \text{maximum layer of the tree}$}
\State {$\var{threshold} \leftarrow \text{the threshold for a node being considered as a child}$}
\Function{ExtendTree}{\var{parent,PList,count}}
\If {$\var{count}>\var{maxLayer}$}
\State {$\var{count}=0$}
\State \Return
\EndIf
\If {$\var{parent}\in\var{PList}$}
\State \Return
\EndIf
\If {$\var{parent} \notin \var{PList}$}
\State{$\var{PList}.add(\var{parent}$)}
\State{$\var{count}\leftarrow \var{count}+1$}
    \For {$i\in (0,n)$}
        \If {$i \notin \var{PList} \ \textbf{and}\  \var{Prob}[\var{parent},i]>\var{threshold}$}
            \State {$\var{childList}.add(i)$}
        \EndIf
    \EndFor
    \For {$\var{child} \in \var{childList}$}
        \State {$\var{ExtendTree}(\var{child, Plist, count})$}
    \EndFor
\EndIf
\EndFunction
\end{algorithmic}
\end{algorithm}

\subsection{Construction of a Bayesian Network}
\tab Dominant features extracted from snapshot matrices are used for obtaining pairwise dependencies. Each snapshot matrix corresponds to a stimulation of a single neuron. We assume the stimulated neuron is driven into active state and thus the snapshot matrix reflects the response of other neurons, which we transform to probabilities, conditioned on stimulated neuron being activate. We achieve this by normalizing each mode according to the response of its input neuron, $X_i$. We interpret the result as the conditional probability $P(X_j=1|X_i=1)$, and store it into the $i^{th}$ column of the conditional probability table. $P(X_j=0|X_i=1)$ can be calculated by $1-P(X_j=1|X_i=1)$, and we assume that $P(X_j=0|X_i=0)=1$ and $P(X_j=1|X_i=0)=0$, i.e. a node cannot be in active state if there is no input to the network.

The resulting table is a $n\times n$ dependency matrix, which records the complete pairwise dependencies for the entire neuronal network. The matrix itself contains useful information about the network, and by constructing a PGM we can further extract and visualize this information. For example, with PGM we can approximate the values of one or more neurons using probabilistic inference, given observations of other neurons.

Since any joint distribution can be decomposed into a product of conditional probabilities, the dependency matrix when transformed into pairwise conditional distributions records the full joint distributions encoded by PGM. A natural graphical representation of a neuronal network is a directed cyclic graph, since the majority of them incorporate multiple pathways and recurrence. Inference on such graphs is hard to perform, as the existence of cycles often leads to non-convergence of the probabilities \cite{Koller}. We therefore choose to eliminate cycles and propose a Restricted Iterative Deepening algorithm (RID) that builds a tree for each posterior inference problem on the graph. The tree structure ensures that each neuron has at most one parent and therefore acyclic by construction. A directed spanning tree can be constructed by choosing an arbitrary root and direct edges away from the root \cite{Koller}. As our goal is to identify functional sub-circuits within the network and their component neurons, tree-structured graph allows us to capture the propagation of neural pathway subject to particular stimuli.

To construct the tree our RID algorithm starts with designated input neuron and extends the tree according to pairwise conditional dependencies. The process continues until the tree either reaches to the desired set of output neurons, e.g., motor neurons, or reaches the maximum depth of the tree, which can be imposed. We restrict the width of the tree (the number of children that a parent can have) by imposing an arbitrary threshold on the conditional dependencies. For each parent neuron $X_i$, we explore all of its children neurons $X_j$ that have pairwise dependencies $P(X_j|X_i)$ exceeding the threshold probability in descending order. 

\section{3-node Neuronal Network Motifs Examples}
\begin{figure}[h!]
  \[\includegraphics[width=14cm]{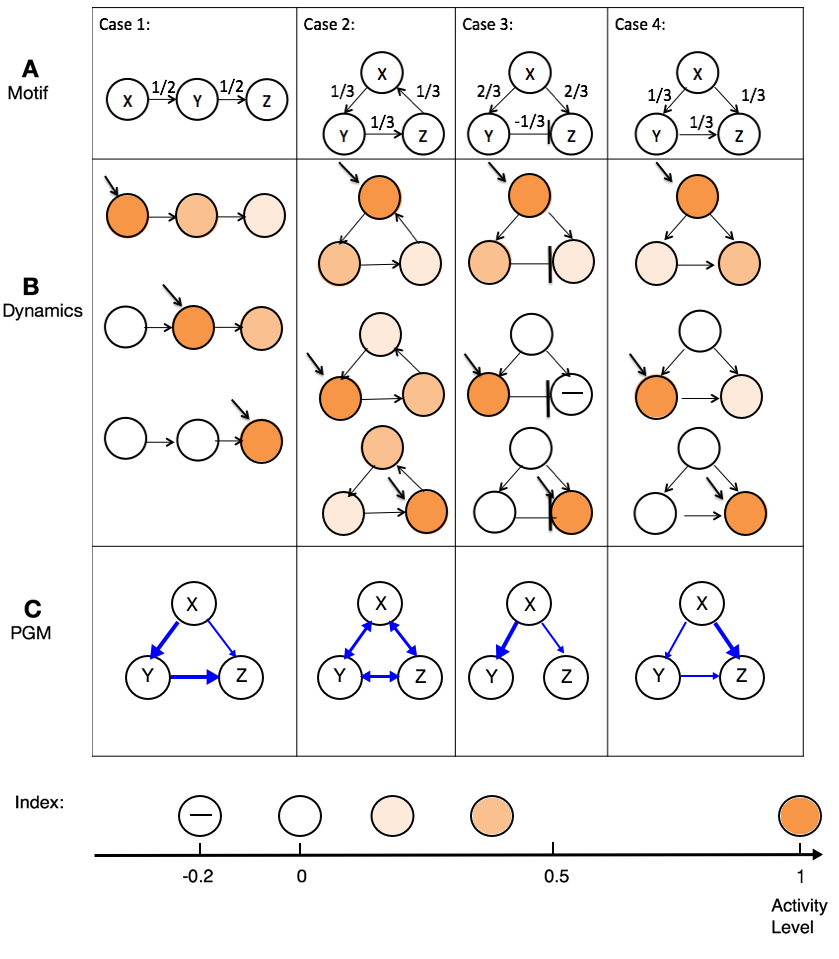}\]
  \caption{\textbf{Construction of PGMs for 3 unit motifs} A: Connectomes of four examined motifs. B: Network responses when external input is injected into each unit (indicated by diagonal arrow). The color of the units indicate the activation level of each unit, i.e., darker color indicates a more active node. C: Constructed PGMs structures. If there is an edge from $X$ to $Y$ , then the conditional probability $P(Y|X)>0.1$. Stronger arrows correspond to higher probability.}\label{fig:motifs}
\end{figure}

\tab To illustrate our methodology we consider example motifs with three units.
Neurons' dynamics are set by a continuous time recurrent neural network (CTRNN) \cite{ctrnn}:
\[ \frac{du_i(t)}{dt}=-\frac{u_i(t)}{\tau_i}+\sum_{j=1}^m w_{ji}\sigma(u_j(t))+I_i(t), \ i=1, \cdots, m\]
where $u_i(t)$ is the internal state of the $i^{th}$ unit, $\tau_i$ is the time constant of the $i^{th}$ unit, $w_{ij}$ are the connectivity signed weights (+ activation, - inhibition), and $I_i(t)$ is the input into $i^{th}$ unit. $\sigma(u_i(t))$ specifies the output of the $i^{th}$ unit, with $\sigma$ being the output function. Here we use $\sigma(x)=\tanh(x)$. We use random variables $X$, $Y$, $Z$ to denote the three nodes in the respective PGM, each takes binary values $\{0,1\}$.

In the case of three units, there are several ways to connect them. We choose four distinct connectivity configurations that provide distinct functionality, shown in Figure \ref{fig:motifs}A. The configurations are: (1) A simple chain from $X$ to $Y$ to $Z$, with weights $w_{XY}=w_{YZ}=\frac{1}{2}$; (2) A simple loop, with weights $w_{XY}=w_{YZ}=w_{ZX}=\frac{1}{3}$; (3) Inhibition edge from $Y$ to $Z$, with weight $w_{YZ}=-\frac{1}{3}$, $w_{XY}=w_{XZ}=\frac{2}{3}$; (4) Three edges that consist a directed acyclic network, with weights $w_{XY}=w_{YZ}=w_{XZ}=\frac{1}{3}$. All other unassigned weights are 0. We call these static connectivity maps as connectomes. Our goal is to infer functional connectivity based on the connectome and the network dynamics.

To simulate the dynamics, we inject a constant input of 1 unit into a specific neuron, i.e. $I_i(t)=1$ and record network response in an appropriate snapshot matrix. We then apply SVD on the snapshot matrix and extract the dominant modes, as we described in the previous section. We repeat this process for each neuron in the network. The probability dependencies are derived from normalizing the dominant modes. For example, in case 2, the network
\begin{align*}
\frac{du_1(t)}{dt}=-u_1+\frac{1}{3}tanh(u_1)+I_1;&
\frac{du_2(t)}{dt}=-u_2+\frac{1}{3}tanh(u_2)+I_2;&
\frac{du_3(t)}{dt}=-u_3+\frac{1}{3}tanh(u_3)+I_3&
\end{align*}
is simulated for a sufficiently long time with stimuli $[I_1, I_2, I_3]$ set to  $[1,0,0], [0,1,0], [0,0,1]$ independently. In all configurations, all nodal responses reach a fixed point. Applying SVD provides the dominant patterns, which correspond to this fixed point, and normalization yields the pairwise dependency matrix $P$
\[P=\bcm 1 & 0.0817 & 0.2507\\ 0.2507 & 1 & 0.0817\\ 0.0817 & 0.2507 &1\ecm\]. The elements of $P$ are interpreted as conditional probabilities
\[P(Y=1|X=1)=0.2507,\ P(Z=1|X=1)=0.0817\]
\[P(X=1|Y=1)=0.2507,\ P(X=1|X=1)=0.0817\]
\[P(X=1|Z=1)=0.2507,\ P(Y=1|Z=1)=0.0817\].

Due to the simplicity of the motifs, we validate the PGM via analytical dynamical systems approach, by calculating the fixed points. For case 2, for example, the fixed point induced by input into node $X$, is $(u_1*,u_2*,u_3*)\approx(1, 0.2507, 0.0817)$. This is exactly the value of the conditional probabilities we computed using  PGM construction approach.
Since all edge weights are identical, same fixed point will be induced by input into the other two nodes up to shuffling of the indices. Similar validation process can be done for other configurations, and we include them in the Appendix section. To infer the functional connectivity graph, we apply the RID algorithm to the dependency matrix $P$ and set the threshold to be $0.1$. Here the maximum depth of the tree is three. Since there are only three nodes we do not restrict the number of children of a parent node. Three individual trees are constructed for each case. Combination of them  into one graph is shown in Figure \ref{fig:motifs}C. Comparing Figure \ref{fig:motifs}A (Connectome) with \ref{fig:motifs}C (PGM) we observe that the PGMs have different structures than their corresponding connectomes. Indeed, the edges in the PGM represent the conditional dependencies instead of weights and reflect how the motif is processing inputs. In particular, in case $1$ the PGM shows that $Y$ strongly depends on $X$, and $Z$ strongly depends on $Y$, which corresponds to the chain structure of the motif. In case $2$, the PGM indicates symmetry and interdependence between all the nodes. Thereby, functionally input injection into any node will produce equivalent response, a characteristic of a circular structure of the motif in this case. Notably, the motif's connectome shows propagation in one direction, however the PGM estimates symmetric behavior. In case $3$, the nodes $Y$ and $Z$ are not connected in the PGM, although there is an edge (inhibitory) between them in the connectome. The PGM construction identifies this connection as non functional since dependency is assigned only in terms of activation. In addition, even though the weights $w_{XY}$ and $w_{XZ}$ are the same in the connectome,  the dependency from $Y$ to $X$ in the PGM is stronger than the dependency from $Z$ to $X$. Such an effect is due to inhibition of stimulus propagation from $X$ to $Z$ through $Y$. In comparison, in case $4$, the connectome has a excitatory edge between $Y$ and $Z$. In this motif, PGM indicates that $Y$ enhances the propagation of stimulus from $X$ to $Z$, even though the weights $w_{XY}$ and $w_{XZ}$ are the same in its motif. These exemplary cases demonstrate that PGM structure is different than the motif's connectome, and captures the functional dependencies between the nodes. These dependencies are not trivial to conclude from the connectome structure alone and become more complex as the dimension of the motif and ratio of connections change.

\section{Application to Neurobiological Dynamic Connectome}
\subsection{Neuronal Network of \textit{Caenorhabditis elegans}}
\begin{figure}[h!]
  \[\includegraphics[width=12cm]{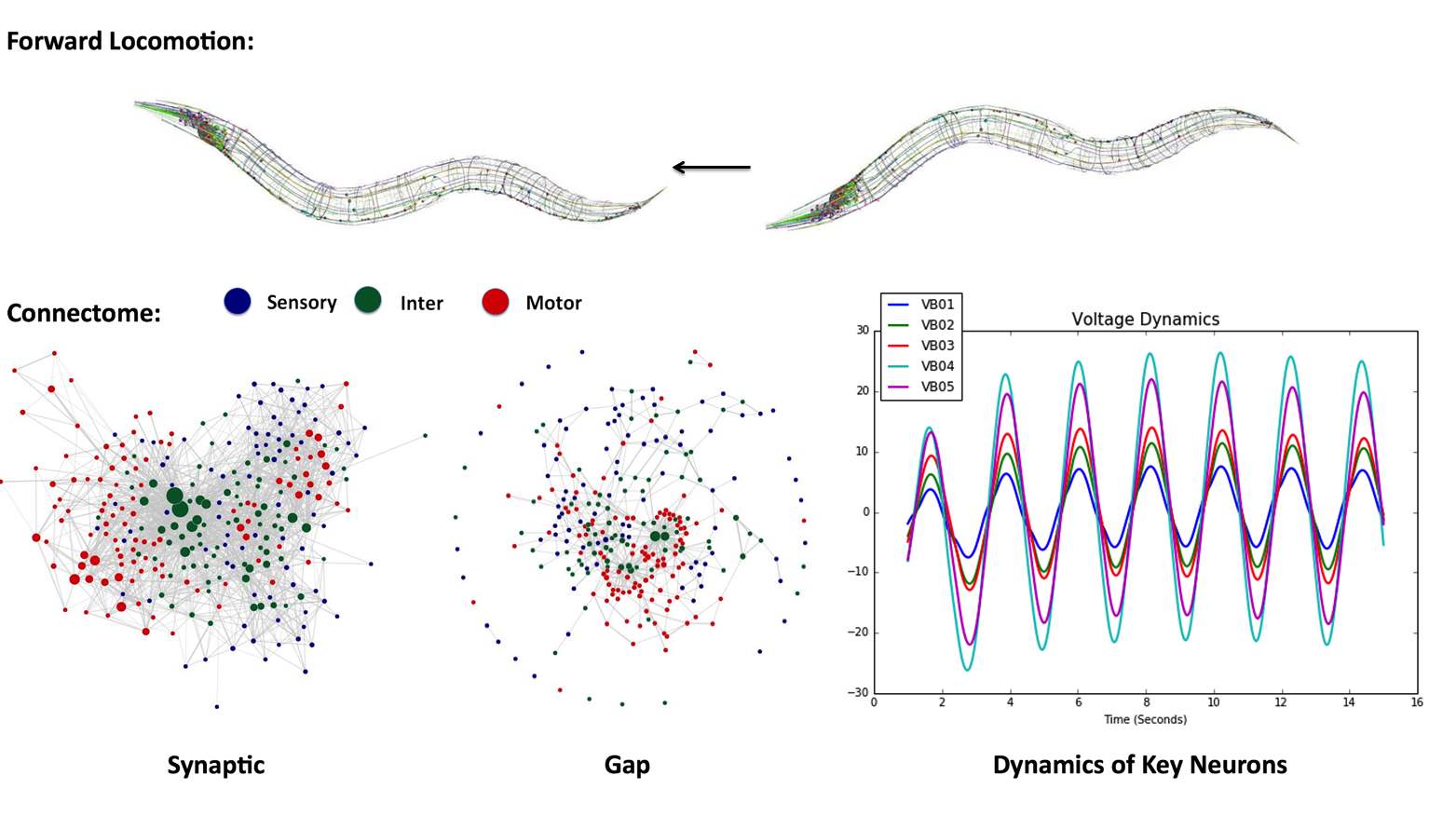}\]
  \caption{\textbf{Two layers of the neuronal network of C. elegans.} Top: An example of forward locomotion induced by two layers of the neuronal network of the worm. The first layer (Bottom Left): Connectome of \textit{C. elegans}, consisting of 297 somatic neurons, 6393 chemical synapses and 890 gap junctions. The left part of the connectome shows the chemical synapses between neurons, and the right part shows the gap junctions. The second layer (Bottom Right): Neural dynamics modeled by differential equations. Here we show voltage oscillations of the motor VB group. Combining the two layers we achieve the dynome, a dynamically evolving network, which is the foundation for constructing the PGM.}
\end{figure}

The nematode \textit{Caenorhabditis elegans} nervous system is a well-studied system, consisting of 302 neurons identifiable and consistent across individuals. The connections between the neurons are comprised of chemical synapses and gap junctions, whose wiring diagrams, i.e., connectomes, are nearly fully resolved from serial section electron microscopy \cite{connectome}. In addition to the connectomes, the dynamic model which describes the biophysical processes between neurons has been introduced \cite{dynome}. Specifically, the dynamical model of the nervous system is governed by a system of nonlinear differential equations:
\[C\dot{V}_i=-G^c(V_i-E_{cell})-I_i^{Gap}(\overrightarrow{\textbf{V}})-I_i^{Syn}(\overrightarrow{\textbf{V}})+I_i^{Ext}\]
where $C$ is the whole-cell membrane capacitance, $G^c$ is the membrane leakage conductance and $E_{cell}$ is the leakage potential. $I_i^{Ext}$ is the external input current injected to the $i^{th}$ neuron. $I_i^{Gap}$ and $I_i^{Syn}$ correspond to the input currents modeling gap junctions and synapses, respectively. More details on the biophysical model can be found in \cite{dynome} and in the Appendix.

Combining the connectome and the dynamical model constitute the dynome model of \textit{C. elegans} network. Incorporating both the layers of connectivity and dynamic biophysical processes, dynome model of \textit{C. elegans} network models the nervous system functionality and stimuli processing that it performs. Indeed, when provided with arbitrary input stimuli, \textit{C. elegans} dynome is capable of producing various forms of characteristic dynamics such as static, oscillatory, non-oscillatory and transient voltage patterns consistent with experimentally observed ones \cite{dynome}. These simulated dynamics indicate that \textit{C. elegans} dynome is a valuable model for the worm's nervous system and thus a suitable foundation for the construction of its probabilistic graphical model.

\subsection{Constructing Dependencies}
\tab We apply our method to the neuronal network of \textit{C. elegans} nematode by injecting scaled input current into each of the $n=279$ neurons independently. We run the simulations for $15$ seconds with time step of $0.01 sec$ and record the dynamics of all neurons in snapshot matrices, with each snapshot matrix $S$ of dimensions $n \times T = 279\times 1501$. We then subtract the activation threshold for each neuron from the simulation, and exclude the initialization phase of the network ($1 ~sec$). Performing SVD on each snapshot matrix and taking the weighted sum of all modes as described in Section \ref{sec:construction}, we obtain $279$ response vector representations of all neurons to the stimulation of the input neurons. Each of these vectors is normalized according to input neuron response, which yields the conditional probability $P(X_j=1|X_i=1)$ as elements of the vector. The vectors are stored as the column vectors of the conditional probability table, resulting in a $279\times 279$ dependency matrix, that records the complete pairwise dependencies of the nodes in \textit{C. elegans} neuronal network.

\section{\textit{C. elegans} Functional Connectome Represented by PGM}
\subsection{Anatomical Connectomes Compared to  PGM Functional Connectome}
\begin{figure}[h]
  \[\includegraphics[width=\textwidth]{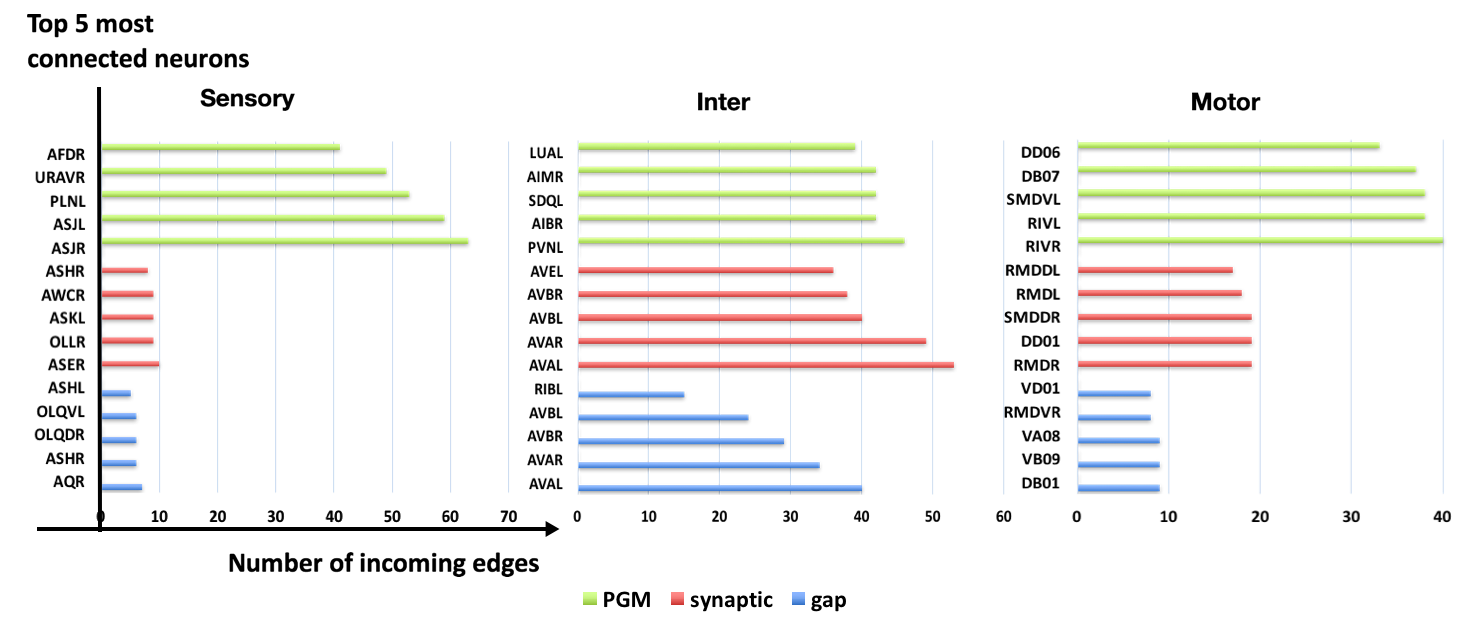}\]
  \caption{\textbf{Top connected neurons in sensory, inter and motor groups.} We compare three graphs (from top to bottom):  functional connectivity represented by PGM, connectome mapping chemical synapses, and a connectome mapping gap junctions,. Connectivity is measured by the number of incoming edges. Top five connected neurons in each graph along with the number of incoming edges into them are shown. In the synaptic graph, an edge exists from one node ($X_i$) to another ($X_j$) if there is at least one chemical synapse $X_i \to X_j$. In the gap graph, an edge exists between $x_i$ and $x_j$ when there is at least one gap junction between $x_i$ and $x_j$, as the gap junctions are non-directed. In the PGM, an edge exists from $x_i$ to $x_j$ if $P(Y=1|X=1)> \textbf{threshold}$, set here as $\textbf{threshold}=0.1$} \label{fig:topconn}
\end{figure}

\begin{figure}[t!]
    \[\includegraphics[width=\textwidth]{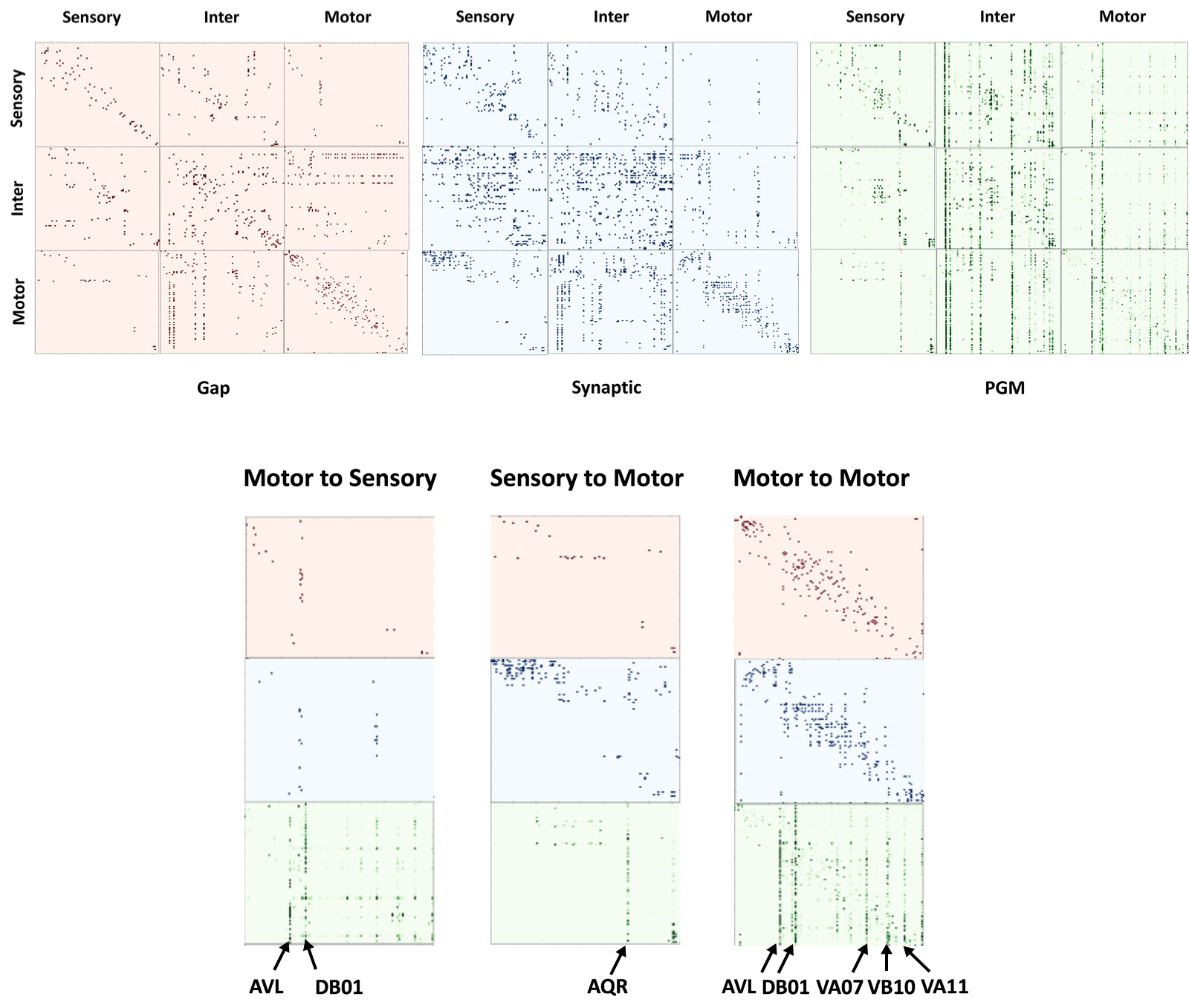}\]
    \caption{\textbf{Connectivity matrices of gap, synaptic and PGM} In the dependency matrices, the element in the $j^{th}$ column and $i^{th}$ row, $a_{ji}$, represents the total number of gap contacts/synaptic contacts, probability dependencies respectively from $i^{th}$ neuron to $j^th$ neuron. The subplots: motor $\rightarrow$ sensory, motor $\rightarrow$ motor, sensory $\rightarrow$ motor are zoomed-in to compare detailed differences in gap, synaptic and PGM connections. Input neurons that activate the majority of the responding neurons are labeled.}\label{fig:connmatrices}
\end{figure}

\tab We compare the dependency matrices (functional connectome) obtained from our PGM construction with the anatomical (gap and chemical connectomes) in Figures \ref{fig:topconn} and \ref{fig:connmatrices}. Figure \ref{fig:topconn} compares top connected (hub) neurons in each group type (sensory, inter, motor) across the three different connectomes. The definition of connectivity for synaptic and gap connectomes is straight forward and we use the number of incoming edges as a count for connectivity. The connectivity in the PGM is expressed through probabilistic interaction and thereby top connected neurons are those having the highest conditional probability. Notably, the PGM identifies vastly different set of hub neurons than those by synaptic and gap connectomes. In particular, we observe that `hub' neurons in the synaptic and gap connectomes, such as AVA and AVB, are not listed in PGM's top connected neurons. Furthermore, top sensory and motor neurons in PGM receive far more connections than those in synaptic and gap connectome.

In sensory neurons, PGM highlights ASJ, PLNL, URAVR and AFDR as neurons with most probabilistic interactions. These neurons are reported to be associated with avoidance behaviors under different circumstances in environment. ASJ neurons take part in light sensation and promote reversals while PLN neurons are part of sensory group associated with oxygen sensing \cite{worm}. URA neurons are generally considered as sensory but also innervate head muscles via the nerve ring. AFD neurons are main thermo-sensors and promote turn under encounters of high temperature. Noting that sensory neurons are often considered as upstream layer of signal pathways, the results show the potential importance of avoidance/turn behaviors in the network.

Inter-neurons associated with avoidance and locomotion appear to be dominating as well. PGM identifies LUA, PVN, SDQ, AIM, AIB neurons as top connected. While the exact functions of LUA and PVN are not very well known, LUA is suggested to be connector cells between PLM touch receptors and ventral cord, suggesting its potential role in locomotion. While AIM neurons are speculated to modulate the locomotion circuit via regulating the extra-synaptic serotonin, SDQ and AIB neurons on the other hand are known to be associated with high oxygen avoidance and promotion of turn respectively \cite{worm}. In motor-neurons group we observe that the most connected neurons are highly associated with turning/locomotion behaviors as well. Both RIV and SMD neurons are known to innervate neck/head muscles that modulate avoidance/escape behaviors such as omega turns, and both DD and DB neurons are main modulators of turns and locomotion in dorsal cord. Taken together, the results provide new insights about the worm's neural function by (i) implying the importance of avoidance/turning behaviors to the organism, and (ii) suggesting potential functional importance of neurons which there roles are unclear. 

In Figure \ref{fig:connmatrices} we visualize the PGM dependency matrix and the anatomic connectomes, with direction of connectivity being $(from,~to)$ side by side, with the neurons ordered by location (anterior to posterior), similarly to the order in \cite{connectome}. This visualization shows the unique characteristics of each connectome. Gap connections appear to be mostly local, i.e., clustered around the diagonal, in which physically neighboring neurons have gap junctions. In addition a few horizontal and vertical ``chains" in inter and motor neurons, which correspond to single neurons having gap junctions with multiple neurons. Synaptic connections incorporate dense connectivity patterns (inter-inter, sensory-inter) in addition to local structure and a few connectivity chains. 
Furthermore, sensory $\leftarrow$ inter and inter $\leftarrow$ motor connections are well-established in the anatomical connectomes, while sensory $\leftarrow$ motor connections are more rare.

PGM connectivity appears to be structured differently than the anatomic ones. Local response is less profound. Most of the dominant patterns appear to be vertical chains, each corresponding to a receiving neuron, triggered by stimulation of multiple neurons across groups, even distant neurons with no gap or synaptic connections. The receiving neurons are from inter and motor groups. In addition, excitation of sensory neurons appears to lead to excitation of motor neurons, and motor neurons reversely also impact sensory neurons. Such observation is consistent with the known ability of sensory neurons to trigger motor behaviors and motor neurons to influence sensation. Another observation from PGM visualization is that there are no dominant horizontal chains, indicating that a single neuron response is triggered by only a few input neurons. According to these observed
properties, we conclude that \textit{neural pathways}, dominant information propagation paths from a neuron of interest or a cluster of neurons should be inferred through posterior inference that will traverse the PGM. 

\subsection{Inference of Functional Sub-circuits}
\tab We perform two posterior inferences on the pairwise conditional table: (i) Given a set of input neurons, what would be the set of downstream neurons most likely to be activated. For example, such an inference is relevant to identify inter and motor neurons activated by a set of sensory neurons. (ii) Given a set of downstream neurons, what are the subsets of upstream and midstream neurons most likely to activate the downstream neurons. Such inference can reveal, for example, inter neurons and sensory neurons that are most likely to activate motor neurons. Both of these questions can be answered by constructing a graphical model for the network and performing posterior inference. While the first problem follow stimulus propagation forward, the second problem is an inverse problem, often called Maximum A Posteriori (MAP) problem, and requires to examine and optimize over many probable inverse propagation sequences. We summarize our key results in below, and include individual neuron trees in the Appendix.

\begin{figure}[t!]
  \[\includegraphics[width=\textwidth]{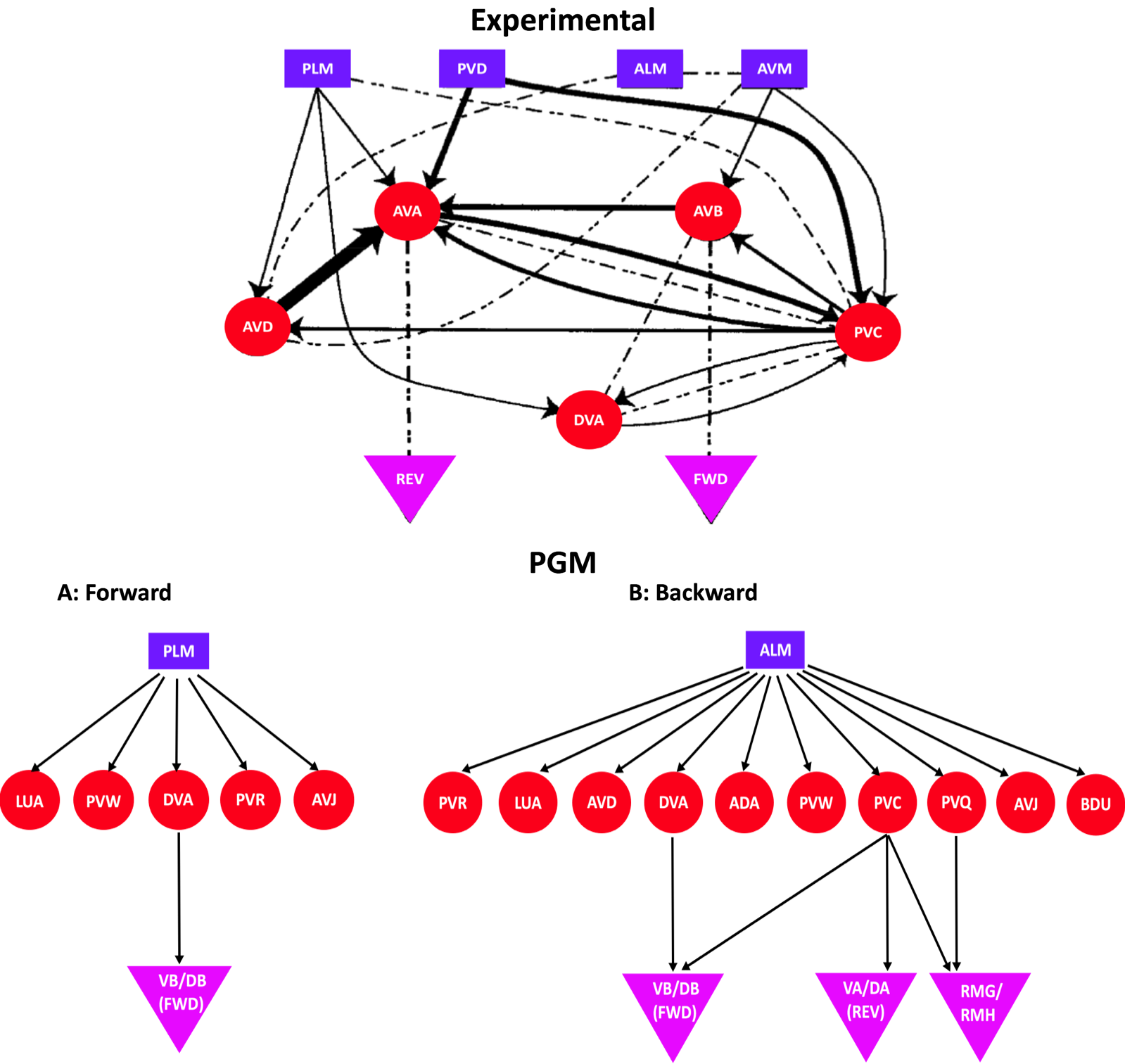}\]
  \caption{\textbf{Sub-circuits that lead to forward and backward locomotion in \textit{C. elegans}} Top: Experimentally proposed connectome sub-circuit of forward and backward locomotion (sensory neurons are blue rectangles, inter neurons are red circles and motor neurons are purple triangles). Lines with arrows refer to chemical synapses and dashed lines refer to gap junctions. Reproduced from \cite{dynamic1}. Bottom: sub-circuits inferred from PGM. A:  Forward locomotion induced by injecting input into sensory neurons PLML and PLMR, known to trigger forward locomotion upon posterior touch. B: Backward locomotion induced by injecting input into ALML and ALMR, known to trigger backward locomotion upon anterior touch.}\label{fig:fwdbwdloc}
\end{figure}

To infer the set of downstream neurons we start with a set of input nodes and use the RID algorithm to construct a response tree for each input node in the set. For each parent node we explore all its children that have conditional dependency exceeding 0.1 in descending order. For simplicity, we restrict the tree to have a maximum depth of 3: one layer of sensory, inter and motor neurons each, to keep the flow from sensory to motor neurons straightforward.

With posterior inference we target well-known experimental scenarios. Specifically, we focus on forward and backward locomotion, as shown in Figure \ref{fig:fwdbwdloc}. In particular, we investigate forward locomotion triggered by posterior touch: activation of sensory PLML/R neurons results in the activation of the DB and VB groups, associated with forward locomotion. We verify this sub-circuit by constructing a tree with input nodes PLML and PLMR (labeled PLM). Both neurons activate the same set of inter neurons, LUA, PVR, PVW, AVJ, DVA, i.e. these inter neurons have the highest probabilities conditioned on the activation of PLML and PLMR sensory neurons. The other inter neurons do not lead to any immediate motor neurons.  DVA neuron leads to DB01 motor neuron in group B (experimentally identified as backward motor neurons), which in turn excites most of the motor neurons in DB and VB groups. These results are consistent with functional stimulus propagation flow reported in literature.  

We also examine anterior touch, triggered by stimulation of ALML and ALMR (labeled ALM) neurons and leads to backward locomotion. ALM activation leads to a larger set of inter neurons. Notably, the set of inter neurons associated with backward locomotion contains all inter neurons associated with forward locomotion and additionally includes AVD, ADA, PVC, PVQ and BDU, most of which indeed associated with backward movement. Indeed, AVD is experimentally known to be associated with backward locomotion, and PVC  plays an important role in both forward and backward motion \cite{worm,dynamic1}. Stimulus flows from PVC neuron to several motor neurons, especially neurons in group A ,i.e., DA, VA (experimentally identified as backward motor neurons) , as well as in DB and VB groups. 

Compared with the circuit extracted from the synaptic and gap connectomes in Figure \ref{fig:fwdbwdloc}Top from \cite{dynamic1}, the PGM successfully recovers neurons participating in this circuit and separates them in two functional sub-circuits. In each of the sub-circuits, key motor neurons are reached to induce locomotion. As can be seen from the circuit sketch, separation of it into independent ones is non trivial. The PGM also identifies  key inter neurons such as AVD, PVC and DVA, although the hub neurons AVA and AVB are missing. A possible explanation is that activation ALM or PLM individually is  insufficient for AVA/AVB to reach a strong state of excitation and they are activated through another stimulation/process.
\begin{figure}[h!]
    \centering
    \includegraphics[width=\textwidth]{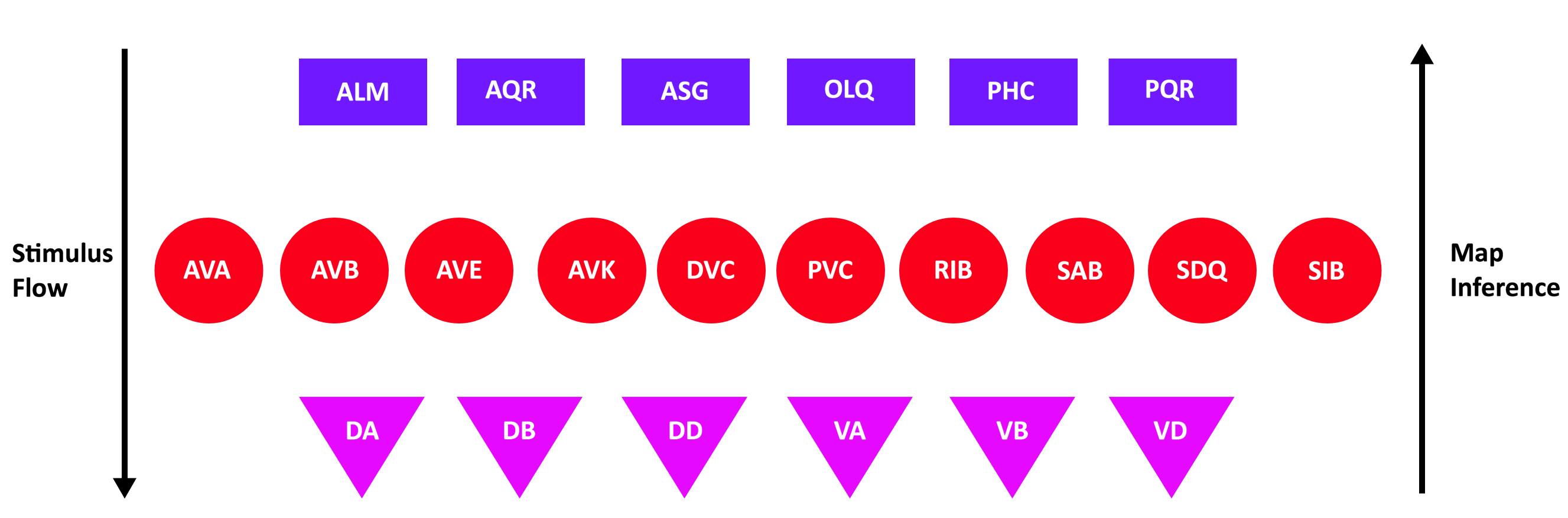}
    \caption{\textbf{MAP posterior inference: Reverse tracking of A and B motor groups associated with locomotion} A group of neurons we identify using MAP inference. Given that a specific group of motor neurons are being excited, the set of inter and sensory neurons that leads to such excitation are shown. From top to bottom is the natural stimulus flow, i.e. sensory $\to$ inter $\to$ motor. From bottom to top is MAP inference. }\label{fig:MAP}
\end{figure}

We use MAP inference to perform reverse tracking of activating neurons of motor neurons associated with locomotion. We, thus, choose motor neurons in the ventral cord members of A and B group shown as purple triangles in Figure~\ref{fig:MAP} and labeled as DA, VA, DB, VB, DD, VD. Posterior MAP inference traverses back the PGM to inter and sensory neurons, which are most likely to activate the given motor neurons. Specifically, we apply the RID algorithm and flipped conditional probabilities. Instead of sorting $P(X_i|X_j)$ we sort $P(X_j|X_i)$, with $X_j$ being the parent node. As in forward RID we limit the  depth of the tree to be 3 and keep one layer of motor, inter and sensory neurons each. We find ten inter neurons and six sensory neurons that most frequently appear in the reverse traversal paths. We list these neurons in Figure~\ref{fig:MAP} as red circles (inter) and blue rectangles (sensory). Notably, most of sensory and inter neurons that MAP inference produces were indeed experimentally associated with locomotion, and these are neurons that we identified in PLM and ALM pathways in Figure \ref{fig:fwdbwdloc}. Furthermore, additional neurons known to participate in locomotion are identified as well. Inferred inter neurons now do include AVA and AVB, which were not present in forward inference from ALM and PLM, see Figure~\ref{fig:fwdbwdloc}, emphasizing that when additional paths are considered, these neurons do play a role in sensory-motor neural integration, but not in direct stimulation of ALM and PLM. The power of MAP inference is in its ability to associate additional neurons with the designated motor neurons, without any prior biological knowledge of the network. For example, additional inferred sensory neurons include AQR and PQR, known experimentally to influence locomotion. Their function is currently being studied and conjectured to be associated with oxygen sensation and avoidance~\cite{chang2006distributed}. Using MAP inference in PGM we, thereby, able to confirm this conjecture. Posterior inference from these neurons would provide more details into which pathways the stimulus from AQR and PQR is following to reach the locomotion motor neurons. 

\section{Discussion}
\tab We presented a new approach to learn a graphical model (PGM) for a neuronal network. Two key components in our approach allow us to construct the PGM that captures network functionality. First is that our underlying component is a dimension reduction technique to obtain a one-dimensional projection of network responses to stimuli. The second is that we capture network responses to independent stimuli (single neuron stimulation). We thus consider pairwise dependencies instead of full dependencies on the whole network, and greatly reduce the number of parameters. This constraint-based approach is computationally efficient, and combined with a method for traversing the PGM to produce response trees (RID) performs posterior inference. We describe how to apply these techniques to simple motif examples and also to the model organism \textit{C. elegans}, which anatomical connectomes and dynamics have been resolved. Application of our method to \textit{C. elegans} dynamical model of neuronal activity identified key sub-circuits without any prior knowledge. Further biological pathways can be extracted from the constructed PGM and the methodology can be applied to other network models and organisms. 

Our framework assumes that the underlying structure of the network remains isotropic over the simulation time, that is, the connectivity of the graph and pairwise conditional probabilities $P(B|A)$ would remain similar if the simulation is continued for longer time. Previous works have suggested that the underlying network could be constantly rewiring due to the long time span of simulation/observation \cite{varying1,varying2}, for example, sequence of networks of gene expression during the life cycle of \textit{Drosophila melanogaster}, which represents different developmental stages. We, thereby, expect our approach to perform best when the simulated network dynamics converge to low dimensional attractors. Indeed, for the network of \textit{C. elegans}, stimuli induced low dimensional attractors were shown to exist in the network and simulation of the network for efficiently long time period ($15~sec$) is expected to capture attractor dynamics \cite{dynome}.

Dimension reduction that we employed is based on Singular Value Decomposition (SVD). We collapse all SVD modes into a single vector by computing the sum weighted by singular values of all modes. Plausible reduction is to retain the modes which energy (singular values) is of $90\%$ of the total energy. In many simulations of \textit{C. elegans}, the first two modes take up more than $90\%$ of the energy, and the first four modes take up more than $99\%$. Indeed, we observe only minor difference between weighted sum of all modes compared to  taking the weighted sum of the first two modes. As expected, we do see a significant difference between using the weighted sum of first two modes and just the dominant mode, indicating that attractors are spanned by at least two dimensional space. Notably, it is also possible to use other approaches to compress the modes into a single vector, e.g., combination of $k-$rank approximation with methods such as Exclusive Threshold Reduction and Optimal Exclusive Threshold Reduction recently introduced in \cite{OETR}.

By independent stimulation and activation of each neuron followed by construction of pairwise response probabilities, our approach assumes that responses could be superimposed, i.e., response probabilities caused by two or more stimuli is the sum of the responses that would have been caused by each stimulus individually. While generally the assumption is not guaranteed to hold in nonlinear neuronal networks, networks which have input induced attractors, as the \textit{C. elegans} system and other attractor systems, many functionalities are an aggregate of response patterns to individual stimulations. Theoretically it is possible to construct a Bayesian network $D(G,\theta)$ with $\theta_i=P(X_i|X_1, \cdots,X_n)$, in which we go beyond the pairwise conditional probabilities and explore all the possible assignments to all the variables in the set. Specifically, for each node in the network, we either activate it, or force it to be inactive. However, doing so requires us to specify $2^n$ distributions, which would require much more simulations and for large $n$ will become computationally intractable.

Our approach for measuring probabilistic dependencies in dynamic processes is different from previously introduced approaches. Alternative approaches propose constructions based on measuring correlations or causality, which collapse time and can possible loose valuable information. Another possibility is to consider time dependent PGMs, however, the posterior inference becomes ill-defined and inefficient as for the original dynamical model from which the PGM is constructed. Our approach is considered as a hybrid of the two aforementioned approaches, since it uses spatio-temporal dimension reduction with pairwise probabilities constraint to produce a static PGM for which posterior inference is fast and well defined.

Posterior inference could be also performed in real-time, by incorporating simulations as the inference occurs, i.e., simulate the network and construct probabilities from time series snapshot matrices. Such an extension will allow to relax the pairwise probabilities constraint and potentially lead to more accurate representation of information flow. We did not choose such an approach since it is generally time consuming and requires performing network simulations for every inference task. In practice, for a network with more than several nodes, real-time inference becomes intractable. We, thereby,  pre-process network responses as pariwise dependencies and construct pre-processed PGM for which inference is a standard inference in a statistical model, which does not involve simulations within posterior inference procedure.

\section{Appendix}
\subsection{Analytic Validation of Motif Examples }
\tab For case 1 a simple chain, we set the weights $w_{XY}=1/2$, $w_{YZ}=1/2$, and all other weights to be zero. Then the network dynamics are:
\[f_X'(t)=-f_X+I_X =f_1\]
\[f_Y'(t)=-f_Y+\frac{1}{2}tanh(f_X)+I_Y=f_2\]
\[f_Z'(t)=-f_Z+\frac{1}{2}tanh(f_Y)+I_Z=f_3\]
\\
1. Fixed point: \[(y_1*,y_2*,y_3*)=(1, 1/2tanh(1), 1/2tanh(1/2tanh(1)))\approx (1, 0.3808, 0.1817)\]
2. Linearization: \[J(y_1*,y_2*,y_3*)=\bcm -1&0&0\\0.5sech^2(y_1*)&-1&0\\0&0.5sech^2(y_2*)&-1\ecm\] 
\[\Lambda=\bcm -1&0&0\\0&-1&0\\0&0&-1\ecm\]
Since all eigenvalues are negative real numbers, the system is stable at the fixed point.\\
Using Algorithm 1, The conditional probabilities obtained are:
\[P(y=1|x=1)=0.3808,\ P(z=1|x=1)=0.1817\]
\[P(z=1|y=1)=0.3808,\ P(x=1|y=1)=0\]
\[P(x=1|z=1)=0,\ P(y=1|z=1)=0\]

For case 3 with feedforward inhibition, suppose $w_{XY}=2/3$, $w_{XZ}=2/3$, $w_{YZ}=-1/3$, and all others are zero. Then the network dynamics are:\\
\[f_X'(t)=-f_X+I_X \]
\[f_Y'(t)=-f_Y+\frac{2}{3}tanh(f_X)+I_Y\]
\[f_Z'(t)=-f_Z+\frac{2}{3}tanh(f_X)-\frac{1}{3}tanh(f_Y)+I_Z\]
\\
1. Fixed point: \[(y_1*,y_2*,y_3*)=(1, 2/3tanh(1), -1/3tanh(2/3tanh(1)))+2/3tanh(1)\approx (1, 0.5077, 0.3517)\]
2. Linearization: \[J(y_1*,y_2*,y_3*)=\bcm -1&0&0\\2/3sech^2(y_1*)&-1&0\\2/3sech^2(y_1*)& -1/3sech^2(y_2*)&-1\ecm\]
\[\Lambda=\bcm -1&0&0\\0&-1&0\\0&0&-1\ecm\]
Since all eigenvalues are negative real numbers, the fixed point is stable.\\
Using Algorithm 1, The conditional probabilities obtained are:
\[P(y=1|x=1)=0.5077, \ P(z=1|x=1)=0.3517\]
\[P(z=1|y=1)=0, \ P(x=1|y=1)=-0.2539 \rightarrow 0\]
\[P(x=1|z=1)=0, \ P(y=1|z=1)=0\]

For case 4 with feedforward excitation, suppose $w_{XY}=1/3$, $w_{XZ}=1/3$, $w_{YZ}=1/3$, and all others are zero. Then the network dynamics are:\\
\[f_X'(t)=-f_X+I_X \]
\[f_Y'(t)=-f_Y+\frac{1}{3}tanh(f_X)+I_Y\]
\[f_Z'(t)=-f_Z+\frac{1}{3}tanh(f_Y)+\frac{1}{3}tanh(f_X)+I_Z\]
\\
1. Fixed point: \[(y_1*,y_2*,y_3*)=(1, 1/3tanh(1), 1/3tanh(1/3tanh(1)))+1/3tanh(1)\approx (1, 0.2539, 0.3367)\]
2. Linearization: \[J(y_1*,y_2*,y_3*)=\bcm -1&0&0\\1/3sech^2(y_1*)&-1&0\\1/3sech^2(y_1*)& 1/3sech^2(y_2*)&-1\ecm\] 
\[\Lambda=\bcm -1&0&0\\0&-1&0\\0&0&-1\ecm\]
Since all eigenvalues are negative real numbers, the fixed point is stable.\\
\\
Using Algorithm 1, The conditional probabilities obtained are:\\
\[P(y=1|x=1)=0.2539, \ P(z=1|x=1)=0.3367\]
\[P(z=1|y=1)=0.2539, \ P(x=1|y=1)=0\]
\[P(x=1|z=1)=0, \ P(y=1|z=1)=0\]

\subsection{\textit{C. elegans} Dynamics}
\tab The dynamic model of the neuronal network is governed by a system of nonlinear differential equations:
\[C\dot{V}_i=-G^c(V_i-E_{cell})-I_i^{Gap}(\overrightarrow{\textbf{V}})-I_i^{Syn}(\overrightarrow{\textbf{V}})+I_i^{Ext}\]
\[I_i^{Gap}=\sum_j G_{ij}^g(V_i-V_j)\]
\[I_i^{Syn}=\sum_j G_{ij}^ss_j(V_i-V_j)\]
where $G_{ij}^g$ is the total conductivity of the gap junctions between $i$ and $j$, and $G_{ij}^s$ is the maximum total conductivity of synapses from $j$ to $i$. $s_i$ is the synaptic activity variable governed by:
\[\dot{s_i}=a_r\phi(v_i; \beta, V_{th})(1-s_i)-a_d s_i\]
where $a_r$ and $a_d$ correspond to the synaptic activity's rise and decay time, and $\phi$ is the sigmoid function with width $\beta$:
\[\phi(v_i; \beta, V_{th})=\frac{1}{1+\exp(-\beta (v_i-V_{th}))}\]

\subsection{Individual Neuron Trees in \textit{C. elegans}}
\tab When constructing individual trees, we extend the depth of the trees to be 5, i.e. for a key sensory neuron we include 2 layers of inter neurons as its children, and for a key inter neurons we include 2 layers of motor neurons as its children. The trees for forward and backward locomotions are obtained by activating ALML/R and PLM/R. Because the resulting trees do not include hub neurons AVA/AVB, we also activate them separately.
\[\includegraphics[width=\textwidth]{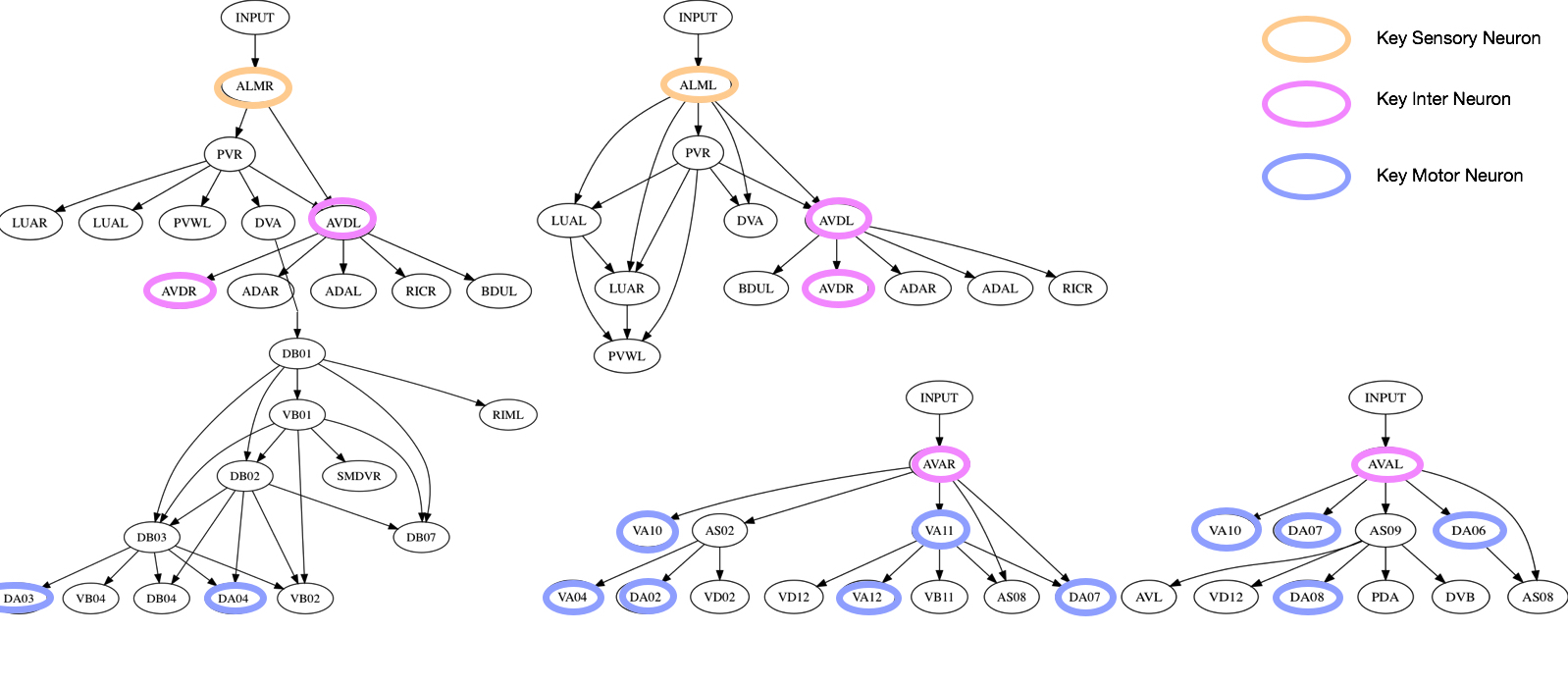}\]
\[\includegraphics[width=\textwidth]{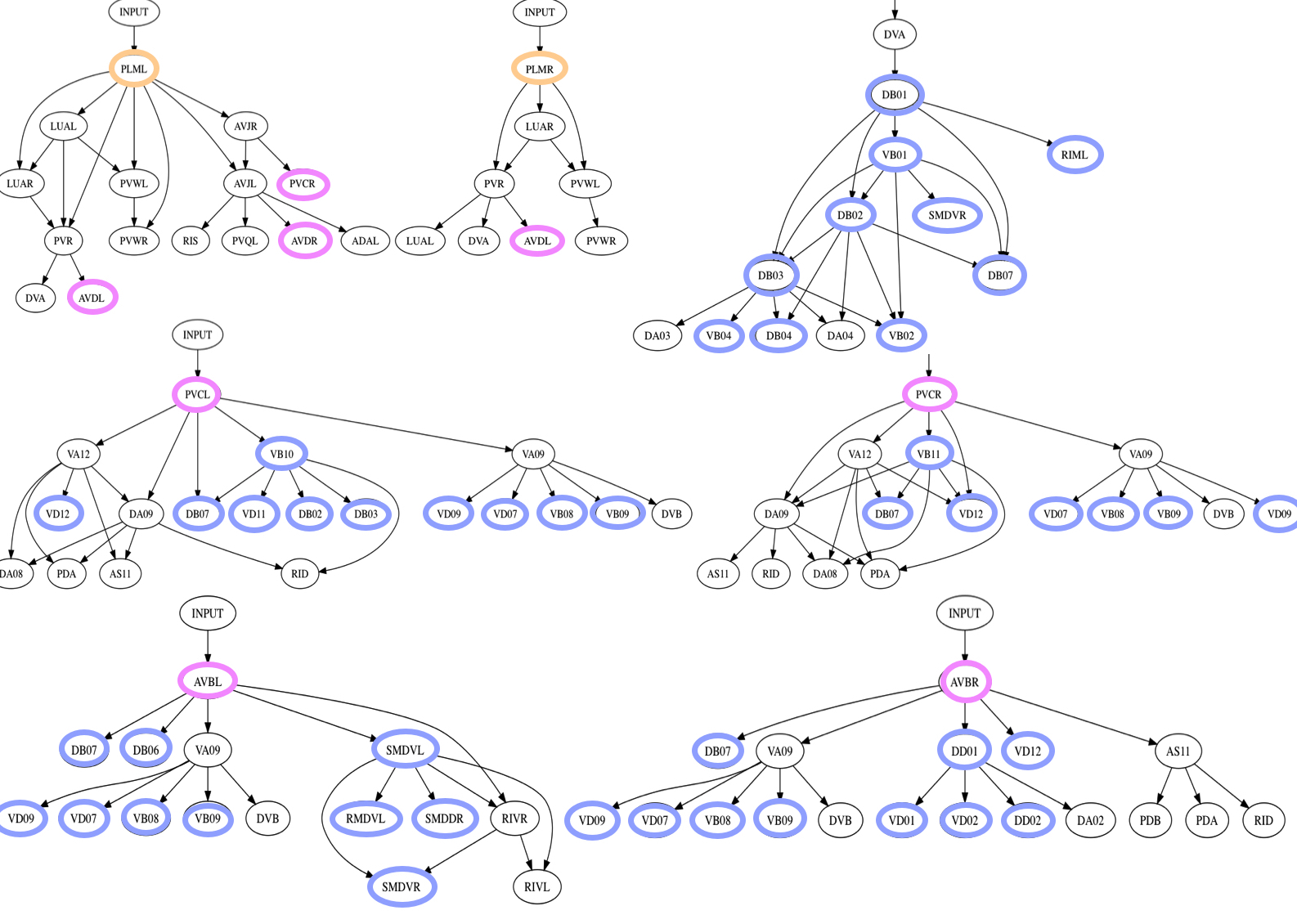}\]

For MAP inference we show the trees starting from the VA groups and the VB groups:
\[\includegraphics[width=\textwidth]{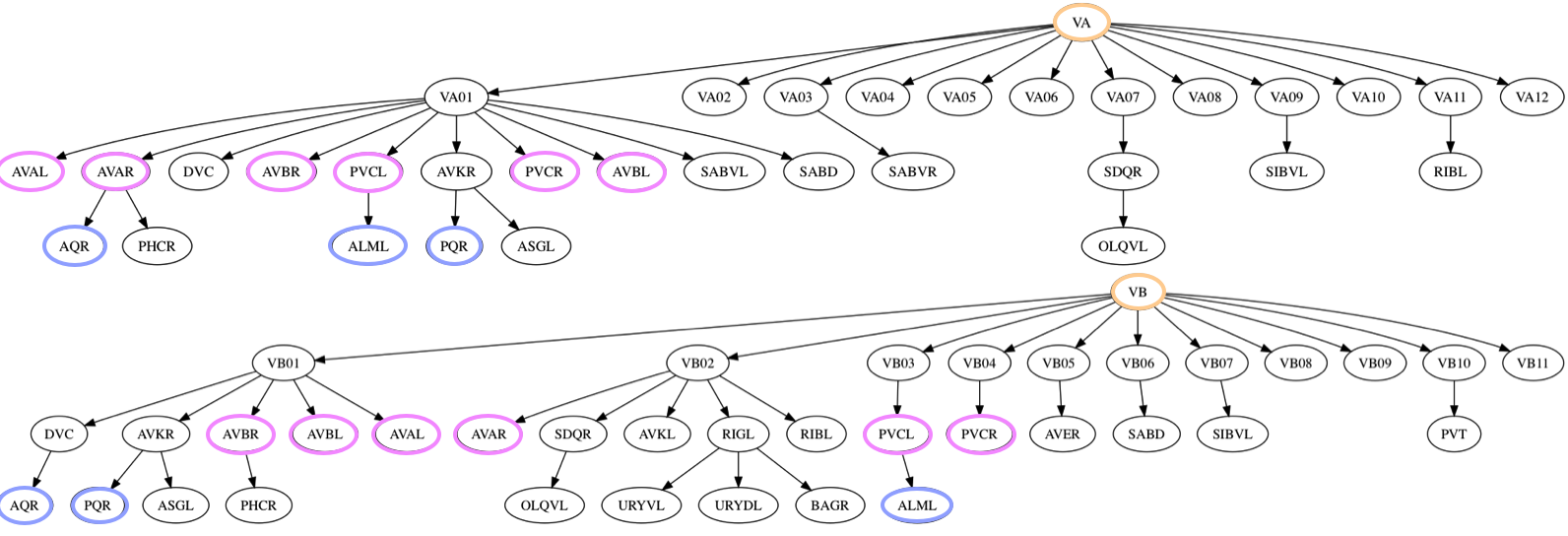}\]

\end{document}